\input iopppt
\input xref
\input epsf
%
%

%
%
\def\allinea#1{%
\null\,\vcenter{\openup1\jot \mathsurround=0pt
\tabskip=0pt\vskip-\smallskipamount\everycr={\noalign{\smallskip}}
\halign{%
&\hfil$\displaystyle{{}##{}}$&$\displaystyle{{}##{}}$\hfil\crcr
#1\crcr
\noalign{\vskip-\smallskipamount}}}\,}
%
\long\def\iniziaitem#1\fineitem
 {{\smallskip\parindent=1truecm
 \parskip=10truept #1\medskip}}
%

\def\ListaFigure{\vfill\eject\global\appendixfalse\textfonts\rm
\everypar{}\null\bigskip}
\def\a{\alpha}
\def\be{\beta}
\def\de{\delta}
\def\g{\gamma}
\def\eps{\varepsilon}
\def\p{\partial}
\def\r{\varrho}
\def\ST{space--time}
\def\CdR{Regge Calculus}
\def\RG{General Relativity}

\def\vphhalf{{\vphantom{\case12}}}
\def\bu{$\bullet$}

\def\vom{v^{\vphantom{j}}_\Omega}

\def\hsmash#1{\setbox0=\hbox{$\scriptstyle#1$} \wd0=0pt \box0}

\def\valass#1{|#1\mkern 1.5mu |}
\def\ugual{\,\mathop=\limits^{\rm def}\,}
\def\tonde#1{$\left(\hbox{#1}\vphhalf\right)$}
\def\tondemat#1{\!\left(#1\vphhalf\right)}
\def\apritonda{$\left(\vphhalf\right.\!$}
\def\chiuditonda{$\!\left.\vphhalf\right)$}
\def\plico{\vrule height 12.7pt width 0pt}

\def\10#1#2{#1\cdot10^{#2}}
\def\funzR{\left(\smash{1-\dot R^2}\vphantom{R^2}\right)}

%
\def\Brewin{Brewin (1987)}

\def\Williams{Collins and Williams (1973)}
\def\Barrett{Barrett \etal (1997)}

\def\Coxeter{Coxeter (1973)}

\def\Sorkin{Sorkin (1975)}
\def\Wald{Wald (1984)}
\def\Recipes{Press \etal (1992)}
\def\nostro{De Felice and Fabri (2000)}
\pptstyle          
\title{Singularities of the closed RW metric in Regge Calculus:
a generalized evolution of the 600-cell}%
[Singularities of RW metric]
\jl{6}

\author{A De Felice%
\footnote{\dag}{e-mail: \tt antonio@astr13pi.difi.unipi.it}
and E Fabri\ddag
\footnote{\S}{e-mail: \tt fabri@df.unipi.it}%
}[A De Felice and E Fabri]

\address{\ddag\ Department of Physics,
University of Pisa, p.zza Torricelli 2, 56126 Pisa PI, Italy}

\abs
An evolution scheme is developed, based on Sorkin algorithm, to evolve
the most complex regular tridimensional polytope, the 600-cell. The
solution of 600-cell, already studied before, is generalized by
allowing a larger number of free variables.  The singularities of
Robertson--Walker (RW) metric are studied and a reason is given why
the evolution of the 600-cell stops when its volume is still far from
zero. A fit of 600-cell's evolution with a continuos metric is studied
by writing a metric generalizing Friedmann's and including the
600-cell evolution too. The result is that the 600-cell meets a
causality-breaking point of \ST.  We also shortly discuss the way
matter is introduced in Regge calculus.
\endabs


\submitted

\date

\section{Introduction}
Regge calculus is 40 years old and has been extensively studied. In its
original form it applies only to an empty spacetime, and the problem of how to
introduce matter has always been a major one.

A key test of \CdR\ in presence of matter which has been studied many
times in literature is the Friedmann universe of dust. The first papers on the
Friedmann universe did not use simplicial complexes. The authors made use of
non simplicial blocks for which they had to give some other information
besides the lengths of the edges. This information was deduced by symmetry.
\Williams\ used three tridimensional regular polytopes to describe the spatial
sections with topology $S^3$ of the Friedmann universe of dust. These authors
assumed $\r$, the dust density, to be constant between two spatial slices.
\Brewin\ wrote in the discrete formalism of \CdR\ the expression of the
action. His expression is equivalent to the one that would result by
assuming a continuous density $\r\propto 1/R^3$.

Besides the regular polytopes, Brewin also used other non regular
tridimensional polytopes. The evolution of all his polytopes, using Regge
equations, stopped before the volume of spatial 3-section became zero: a
point was reached where the equations had no solution. He saw that this point
approaches the limit value (spatial section of null radius) as the polytope
approaches the 3-sphere of the continuum solution. To overcome the obstacle,
Brewin continued the evolution reverting to differential equations.

The first authors who used a 4-dimensional simplicial complex for the
Friedmann universe of dust were \Barrett. They evolved a spatial section made
of a regular 3-polytope having $n$ vertices, and ``discretized'' matter by
placing one mass point on each timelike edge joining a vertex
$i$ of the polytope with its evolute $i'$ of the next spatial section. Thus they
found the action:
$$
S=\frac 1{8\pi }\sum_b\eps _b\,A_b-\frac Mn\sum_il_{ii'},\label{azione}$$
where the first sum is over the bones $b$ of the manifold and $\eps_b$ is the
defect of the bone $b$ with area $A_b$. The second sum is done over every
vertex $i$ of the first spatial section. For a regular polytope having
$n=120$, a mass $M/120$ is put in each vertex $i$.

The Regge equations are found by extremizing the action \ref{azione}:
$$\sum_k\eps _{ijk}\,\frac{\partial A_{ijk}}{%
\partial l_{ij}}=\frac \pi {15}\,M\,\delta_{i'j},$$
where the sum is done over all vertices $k$ joined to the edge $[ij]$
of length $l_{ij}$. The bone $[ijk]$ has $\eps_{ijk}$ as defect and
$A_{ijk}$ as area. They made use of the evolutive scheme found by
\Sorkin\ by assuming that the vertices of the 600-cell could be
subdivided in four classes each composed by vertices not joined
together.

We have found that the vertices of the 600-cell actually belong to
five classes and have investigated if this difference might lead to
different results.

All evolutive schemes described above exhibit a common behaviour: before
reaching the ``true'' singularity of Friedmann universe, the evolution stops
because Regge equations have no solution. It is not clear if this happens
because of the discretization intrinsic in Regge calculus, or for other
reasons (see section \ref{problema}). We decided to investigate this point in
greater detail.

This paper is organized as follows. In section \ref{evol} we sketch
the results of our calculations. In section \ref{secEmb} we embed the
closed Robertson--Walker metric into a five-dimensional Lorentzian
space. We find it useful in order to understand the problems related
to the evolution of our lattice.  The nature of the singularities
belonging to the closed Robertson--Walker metric is discussed in
section~\ref{analisi}. In section~\ref{problema} we propose a deeper
reason why our simplicial approximation of the Friedmann universe
ceases to exist before the volume of the spatial 3-section
vanishes. By fitting the simplicial solution with a continuous metric
(which has the same behaviour with respect to the reaching of
singularities) we find that the stop condition is caused by a novel
singularity of the metric (section~\ref{nuovo}). Finally, our
conclusions are reported in section~\ref{conclusioni}.

\section{Results of the evolution\label{evol}}
The vertices of the 600-cell belong to five classes each including only
vertices not joined together. Using the classification given by \Coxeter,
one of these five classes is composed by the following vertices:
$$\eqalign{%
\alpha&=\{A_0,A_{10},A_{20},A_{30},A_{40},A_{50},
B_{7},B_{17},B_{27},B_{37},B_{47},B_{57},\cr
&\qquad
C_{1},C_{11},C_{21},C_{31},C_{41},C_{51},
D_{8},D_{18},D_{28},D_{38},D_{48},D_{58}\}.\cr}$$
The other classes $\be$, $\g$, $\de$, $\eps$ are obtained by adding 2, 4, 6, 8
to the subscripts.

It should be noted that every evolution consists in solving 5 systems of 4
equations in 4 unknowns each, given the lapse and shift freedom. Furthermore
the simplicial sandwich does not have the same grade of symmetry as the
initial spatial section, where all vertices were equivalent. So the Sorkin
evolutive algorithm we used breaks the symmetry among the vertices $\a$,
$\be$, $\g$, $\de$, $\eps$.

We have generalized the evolution of the 600-cell by imposing that the
edges of the same type, like $[\a\be]$, were all equal. In this way only
four unknowns and four equations remained when each vertex was evolved. By
allowing the edges joining vertices belonging to two different classes to be
possibly different one from another, we introduce new degrees of freedom in
the lattice. We have verified that this enhancement of freedom causes no
problem in the evolution.

We performed the evolution of a 600-cell starting from the instant of time
symmetry. Our results are in figure~\ref{rfunzionet}. The evolution stops at a
finite value $R_{\rm m}$ of the radius $R$ (the radius of the three-sphere
equivalent to the 600-cell) which is approximately 1/4 of the initial
value ($R_0=4.244\dots$). For $R<R_{\rm m}$ the Regge equations give no
solution. Further numerical details may be found in \nostro.
%

\section{The embedding 5-space\label{secEmb}}
We shall find convenient to embed a \ST\ with closed Robertson--Walker
metric into a 5-dimensional Lorentz space. This will be of help for
understanding the behaviour of the Regge equations when applied to the
evolution of a 600-cell space section.

The metric under consideration can be written in this way:
$$
\d s^2 = -\d t^2 +
R^2\left[\d\chi^2+\sin^2\chi\left(\d\vartheta^2+\sin^2\vartheta\,\d\varphi^2
\right)\right],\label{labellag}$$
where $R$ depends only on the universal time $t$.

The manifold with metric \ref{labellag} can be embedded into a 5-dimensional
space endowed with the (1+4) metric
$$\d \sigma^2 =-\d v^2+\d w^2+\d x^2+\d y^2+\d z^2. \label{dsigma}$$
In the following we will refer to $v$ as ``outer time.''

The metric \ref{dsigma} can be rewritten using coordinates $v$, $R$, $\chi$, $\vartheta$, $\varphi$
by the substitution
$$\allinea{%
 w&=R\cos\chi\cr
 x&=R\sin\chi\sin\vartheta\cos\varphi\cr
 y&=R\sin\chi\sin\vartheta\sin\varphi\cr
 z&=R\sin\chi\cos\vartheta.\cr}\label{coordinate}$$
Then
$$
\d\sigma^2 =-\d v^2+\d R^2+
R^2\left[\d\chi^2+\sin^2\chi\left(\d\vartheta^2+\sin^2\vartheta\,\d\varphi^2
\right)\right].\label{gimmersa}$$
A 4-dimensional submanifold of \ref{gimmersa} can be assigned by
imposing that $R$ is a function only of the outer time $v$: then we have
($\dot R\equiv\d R/\d v$)
$$
\d s^2=-\funzR\d v^2+
R^2\left[\d\chi^2+\sin^2\chi\left(\d\vartheta^2+\sin^2\vartheta\,\d\varphi^2
\right)\right].\label{newg}$$
In order that equations \ref{labellag} and \ref{newg} may coincide it is
necessary that
$$\d t^2=\funzR\d v^2,\label{dv}$$
from which it follows that $\dot R^2<1$. We are interested in the case where
$R(v)$ is an even function, so that $v=0$ is an instant of time symmetry.
Integrating equation \ref{dv}, $t$ can be found as a function of $v$.

\section{Singularities\label{analisi}}

In order to study the singularities of metric \ref{newg} let us compute
the quadratic Riemann invariant:
$$Q = R^{\a\be\g\de}\,R_{\a\be\g\de}=
\frac{12}{R^2\bigl(1-\dot R^2\bigr)^2\plico}
\left[\frac{\ddot R^2}{\bigl(1-\dot R^2\bigr)^2\plico}+\frac1{R^2}\right].
\label{riem}$$
We will assume as a necessary and sufficient condition for a singularity that
$Q$ becomes infinite. A glance to equation \ref{riem} shows that the singular
points are those values of $v$ where $\valass{\dot R}=1$ or $R=0$ or
$\valass{\ddot R}=+\infty$. Let us discuss the possible cases.

\iniziaitem
\item{1.}{\sl The points\/} $R=0$. These coincide with the infinite
contraction of the universe when the volume of the spatial section goes to 0.
In these points the density reaches infinite values. By a volume-vanishing
(VV) singularity we will mean one of these points.

\item{2.}{\sl The points\/} $\dot R=\pm1$. If we are not in the first case we
can say that the radius of the space section increases or decreases with speed
tending to that of light. If one tries to extend the manifold beyond
these points, and if $\valass{\dot R}$ increases further, the metric becomes
positive defined. So it can no longer represent a manifold of \RG. This is
tantamount to saying that the spacelike sections are not causally connected:
the points of two different spatial sections can be joined only by spacelike
geodesics. These points will be said to be causality-breaking (CB)
singularities. Actually, in a rigorous sense a CB singularity is a border of
the manifold, and the extension is meaningless.

\item{3.}{\sl The points\/} $\ddot R=\pm\infty$. If we are not in the previous
two cases the singular behaviour is shown by the geodesics deviation, which
becomes infinite. We will refer to these points as geodesics-deviation (GD)
singularities.  \par\fineitem

\noindent It should be noted that in general the GD points do not represent a
singularity according to the definition used in the context of Hawking's
theorem, see for example \Wald, where the presence of a singularity means that
the universe is geodesically incomplete. An example of a GD point which does
not stop any geodesic is the point $v=0$ for the universe having
$R(v)=R_0+A\,v^{4/3}$.

These singularities can also occur together in the same point: an
example is the hypothetical universe whose radius has the form
$R(v)=v-A\,v^{3/2}$ (the critical point $v=0$ is a VV-CB-GD singularity).
Another interesting case of singularity merging is the Friedmann metric
discussed in subsection \ref{friedmetr} below.

\subsection{Matter-limited singularities}
From Einstein equations for the closed Robertson--Walker metric we have for
the diagonal components of $T_{\a\be}$ in an orthonormal basis:
$$\eqalign{%
\r&=\frac3{8\pi}\frac1{R^2\bigl(1-\dot R^2\bigr)\plico}\cr
p&=-\frac1{8\pi}\frac{2R\ddot R+1-\dot R^2}
                     {R^2\bigl(1-\dot R^2\bigr)^2\plico}.}\label{rhop}$$
It is well known that the stress-energy tensor must satisfy
$$\r \geq |p|.$$
In our case
$$-2\le\frac{R\ddot R}{1-\dot R^2\plico}\le1.
\label{rhogeqp}$$
In general it can happen that condition \ref{rhogeqp} be not satisfied for
values of $v$ for which the metric still has a ``geometrical sense.'' In fact
a CB or GD or CB-GD singularity does not satisfy equation \ref{rhogeqp}.
Instead a VV point presents no problem. It should be noted that we cannot
become aware of these new ``physical'' matter-limited (ML) points, wherein
$\r=|p|$, by looking only at the expression of the metric.

\subsection{Friedmann metric\label{friedmetr}}
Since $\d t=R\,\d\eta$, from equation \ref{dv} we obtain
$$
v=\int\!\left[R^2-(\d R/\d\eta)^2\right]^{\!1/2}\d\eta.$$
For the Friedmann metric we have:
$$v=2R_0\,\sin\frac\eta2$$
and
$$R=R_0\cos^2\frac\eta2=R_0-\frac{v^2}{4R_0}.$$
Metric \ref{newg} takes the following form:
$$
\d s^2=-\frac R{R_{\hsmash0}}\>\d v^2+
R^2\left[\d\chi^2+\sin^2\chi\left(\d\vartheta^2+\sin^2\vartheta\,\d\varphi^2
\right)\right].$$

Calling $v_c$ the value of $v$ where $\dot R=-1$ and $\vom$ the positive value
where $R=0$, for the Friedmann metric these two singularities occur
{\sl in the same point\/}, i.e.~$\vom=v_c=2R_0$. So, in the 5-space, in
correspondence of the big crunch the speed of contraction equals the speed of
light, i.e.\ the big crunch is a VV-CB point. This behaviour is shared by all
matter satisfying $p=w\,\r$, where $w$ is a constant, $-1/3<w\le1$, since
then $\dot R^2 = 1 - A^2\,R^{1+3w}$, $A$ a constant. Seeing that $\ddot R =
-\frac12\, A^2\, (1+3w)\,R^{3w}$, in Zel'dovich's interval, $0\le w\le1$, no
GD singularity is present besides the other two.

\section{The problem of the 600-cell\label{problema}}
What is the nature of the critical point for the 600-cell model? Does it reach
a singularity, and in this case which of them? \Brewin\ gives a partial answer
to these questions by noting that the following relation occurs:
$$\frac{\Delta l}\tau\approx1,$$
where $\tau$ is the interval of universal time between two consecutive spatial
sections and $\Delta l$ is the difference of the lengths of their edges.
The collapse becomes so fast that the vertical edges are to become spacelike.
Actually in our model the above fraction exceeds 1 and reaches a not easily
interpretable value (approximately 2.6). \Barrett\ saw the
position of the stop point is quite independent of the timelike interval
between two consecutive spatial sections and advanced the idea that
the evolution should be continued by making spacelike the edges of the lattice
which were timelike before the stop. However they did not pursue the idea. So
what is the true meaning of the above condition, and how can we continue the
iterations (if possible)?

In order to answer all these questions let us now reconsider equation
\ref{dv}. For each evolutive step we know $\Delta t$ ($=\tau$) and $\Delta R$.
Then we can find an approximate value of $\Delta v$ by calculating $\Delta v =
\tondemat{\Delta t^2+\Delta R^2}^{\!1/2}$ and compute $R$ as a function of $v$
(figure~\ref{rfunzionev}).
%
%
We also give $\Delta R/\Delta v\approx\d R/\d v$
as a function of $R$ in figure~\ref{dvfunzionedr}.
%
%
We can see that this derivative in the last iterations tends, in absolute
value, to the unity. In this sense the 600-cell meets a CB singularity before
its volume vanishes. So we expect that for the 600-cell $v_c<\vom$. This leads
us to believe that the evolution cannot be continued beyond this point.

We can get a clearer understanding of the situation examining a case easier to
be studied in detail. Consider the following partial differential equation:
$$\frac{\p y}{\p x}+\frac{\p y}{\p t}=-t-x\,t-\case12\,x^2,
\qquad{\rm with} \qquad y(x,0)=\case12\,.\label{esempio}$$
Its solution is
$$y=\case12-\case12\,t^2-\case12\,t\,x^2.\label{esatta}$$
We can see from equation~\ref{esatta} that the relation $y=0$
is satisfied along two curves. There are also two curves (actually two
parabolas) where $\p y/\p t=\pm1$. These four curves are tangent two by two
in two points, $x=0$ and $t=\pm1$, wherein both conditions hold.

Let us try to solving equation \ref{esempio} by using a numerical
approximation. If we use Lax method \apritonda see \Recipes\chiuditonda,
taking $h$ as the spatial $x$-step and $k$ the $t$-step ($h>k$ for the
Courant condition), the above differential equation becomes
$$\frac{y_{m+1,n}-y_{m-1,n}}{2h}+ \frac{2y_{m,n+1} - (y_{m+1,n} +
y_{m-1,n})}{2k}= -t_n-x_m\,t_n-\case12\,x_m^2.$$
The solution can be easily worked out and we can see that the new four curves
are not tangent any longer. Furthermore the region where $|\p y/\p t|\leq1$
is strictly included into the region $y\geq0$. So the two points in $x=0$ are
splitted and for every $x$ the solution starts and ends in two ``CB points''
($\p y/\p t=\pm1$). This discrepancy vanishes as $h$, the $x$-step,
reaches smaller and smaller values.

The above argument shows that our 600-cell model does not stop because it is
going to reach the singularity $R=0$ ---a situation which, like in all
numerical methods, could give rise to problems. Instead our model stops
because it fits a universe that
reaches another kind of singularity ($\dot R=-1$)

\section{Study of a ``new'' class of solutions of Einstein equations
\label{nuovo}}
We have just seen that the behaviour of the evolution for the 600-cell looks
somewhat different from the Friedmann solution, in its reaching a
CB singularity before the vanishing of $R$. So it appears expedient to
study a generalisation of Friedmann metric, also having that property.
For the 600-cell a good empirical fit of $R(v)$ is given
by %
$$R(v)=R_0-\frac{a^2v^2}{4R_0},\qquad{\rm with}\qquad a^2\approx1.128,$$
whereas $a=1$ is the Friedmann metric.

\noindent It is useful to re-define
$$\allinea{%
&v_c&&\ugual\frac{2R_0}{a^2},\cr
&\vom&&\ugual\frac{2R_0}a=a\,v_c,\cr}$$
so
$$\eqalign{%
R&=\frac1{2v_c}\left(v_\Omega^2-v^2\right)\cr
\dot R&=-\frac v{v_{\hsmash c}}\cr
\ddot R&=-\frac1{v_{\hsmash c}}\>.\cr}$$

\iniziaitem
\item{\bu}For $0<a<1$ we have that $-\vom<v<\vom$ and the singularity is of
VV-type.

\item{\bu}For $a>1$ we have that $-v_c<v<v_c$ and the singularity is of
CB-type. This is the case our model fits.

\item{\bu}For $a=1$ the solution is the Friedmann metric. So this case
is a watershed between two different behaviours of the general metric.
\par\fineitem

\noindent It is not difficult to find the expression of~$t$ as a function
of~$v$:
$$t=\case12\,v_c\arcsin(v/v_c)+\case12\,v\left[1-(v/v_c)^2\right]^{\!1/2}$$
and to show that the graph of~$R(t)$ is still a cycloid, like for Friedmann
metric, but scaled in~$R$ by a factor~$1/a^2$ and translated
upwards by~$\frac12\, v_c\left(a^2-1\right)$.

\subsection{Behaviour of the metric}
Condition \ref{rhogeqp} implies that the following inequality must
be satisfied:
$$v^2\leq\tilde v^2\ugual v_c^2-\case13\,v_c^2\left|a^2-1\right|.$$
If $a<1$ this condition causes no problem because $\tilde v^2 >
v_\Omega^2$. If $a>1$ the ML point $\tilde v$ is the first one the metric
meets in its evolution. In fact
$$\tilde v^2=v_c^2-\case13\,v_c^2\left(a^2-1\right)<v_c^2.$$
Note that in order to have $\tilde v^2\geq0$ the value of $a^2$ must not
exceed 4. So the conditions for the stress-energy tensor reduce the interval
of definition of $a$ to $0<a\leq2$. If $a<1$, the metric ends in a VV
singularity reached the more softly the smaller the value of $a$. If $a>1$,
the metric reaches a ML point before the CB one.
The Friedmann metric ($a=1$) is more ``pathological'' because its VV
singularity is also a CB point, since $\vom=v_c$.

For this metric it is also possible in principle to write down an equation
of state of matter
$$\r=\frac32\left[p^2+\frac4{\sqrt{2\pi}\,v_c}
     \left(\frac p{a^2-1}\right)^{\!3/2}\,\right]^{\!1/2}-\frac32\,p,$$
but of course we do not attach any physical meaning to this result.

\section{Conclusions\label{conclusioni}}
We have seen that the evolution of the 600-cell does not describe the
Friedmann universe well. Instead we can think of it as the evolution of a
different type of matter. But where does such a difference come from? In our
opinion, one should consider the following three points as playing an
important role in answering this question.

First, the 600-cell is a coarse approximation to a 3-sphere.
For every approximating method, \CdR\ included, the smaller the
step interval the better the fit. When this interval reduces, the two
solutions (numerical and analytical) tend to coincide. In \CdR, in order
to get a better approximation, one should increase the number of tetrahedra of
each space section, i.e.\ the 600-cell should be substituted by another
non-regular polytope nearer to a 3-sphere \tonde{\Brewin}.

The second aspect is deeper since it is closely related to the grounds of the
\CdR\ itself. We have seen that Regge equations give rise to a different
evolution of the universe, ending in a CB point. This happens because they are
only an approximation of Einstein equations.

Let us discuss this point by analogy with a related topic. In solving
differential equations it is usual that a numerical method works better than
another. For example let us take the following equation
$$\frac{\d x}{\d t}=f(x,t)$$
and try to find a solution. If we used Euler's method we should write
$$\frac{x_{k+1}-x_k}h\approx\frac{\d x}{\d t}(t_k)+\case12\,h\,
\frac{\d^2x}{\d t^2}(t_k)+\dots=\frac{\d x}{\d t}(t_k)+O(h),$$
whereas, using the central-differences method, we would obtain
$$\frac{x_{k+1}-x_{k-1}}{2h}\approx \frac{\d x}{\d t}(t_k)  +
\case16\,h^2\,\frac{\d^3x}{\d t^3}(t_k)+\dots=
\frac{\d x}{\d t}(t_k)+O\!\left(h^2\right).$$
So, even if both methods lead to the correct solution when $h$ approaches
to~0, the central-differences method is generally more accurate and converges
more quickly than Euler's. According to the approximating method we choose,
we have a smaller or greater truncation error. Furthermore, given that an
approximate value for $\d x/\d t$ implies the presence of other additional
derivative terms of order greater than one, solving a differential equation
through a numerical method is more like to solve another differential
equation, which, as we have seen in section~\ref{problema}, of course has
different solutions which may behave qualitatively different from the one we
are looking for: for instance, the approximate solution may exhibit
singularities not belonging to the correct solution.

The third point is related to the way matter is taken into account.
We followed \Barrett\ by putting each mass point along the vertical
edges, i.e.\ the ``rest'' geodesics that pass through the vertices
of the simplicial complex.  This appears to be the only easy way to
introduce matter in a simplicial complex; so doing, however, a further
discretization is introduced, not logically related to the original
Regge idea. Alternatively, the mass points could be placed along other
``rest'' geodesics; then the proper lengths would be a function not
only of $\tau$ but also of other edges belonging to the region between
two spatial sections. In this case we should obtain a novel expression
for the action and of course new Regge equations which might behave
differently. Furthermore, we could put more than one single mass point
in each tetrahedron, or even think of tetrahedra uniformly filled with
dust.  It is not clear how this change could reflect on the behaviour
of the model.

The undesired halting of the iteration scheme appears to be a property
of all approaches attempted thus far to the Friedmann universe of dust,
irrespective of their different treatment of matter. It is then reasonable to
ascribe such behaviour either to the spatial discretization or to the
Regge equations themselves, as explained before. Our opinion --- admittedly
not entirely justifiable --- inclines towards the latter.

\references
\refjl{Barrett J W 1986}{\CQG}{3}{203--6}

\refjl{Barrett J W, Galassi M, Miller W A, Sorkin R D,
Tuckey P A and Williams R M 1997}{Int. J. Theor. Phys.}{36}{815--39}

\refjl{Brewin L C 1987}{\CQG}{4}{899--928}

\refjl{Brewin L C and Gentle A P 2001}{\CQG}{18}{517--26}

\refjl{Collins P A and Williams R M 1973}{\PR\ {\rm D}}{7}{965--71}

\refbk{Coxeter H S M 1973}{Regular Polytopes}{(New York: Dover
Publications, Inc.) pp~247--50}

\refjl{De Felice A and Fabri E 2000}{{\rm gr-qc/0009093}}{}{}

\refjl{Gentle A P and Miller W A 1998}{\CQG}{15}{389--405}

\refbk{Press P H, Teukolsky S A, Vetterling W T and Flannery B P
1992}{Numerical Recipes in C}{(Cambridge University Press) pp~834--9}

\refjl{Sorkin R D 1975}{\PR\ {\rm D}}{12}{385--96}

\refjl{Tuckey P A 1993}{\CQG}{10}{L109--13}

\refbk{Wald R M 1984}{General Relativity}{(Chicago, University of Chicago
Press)}

\ListaFigure

\centerline{\epsfbox{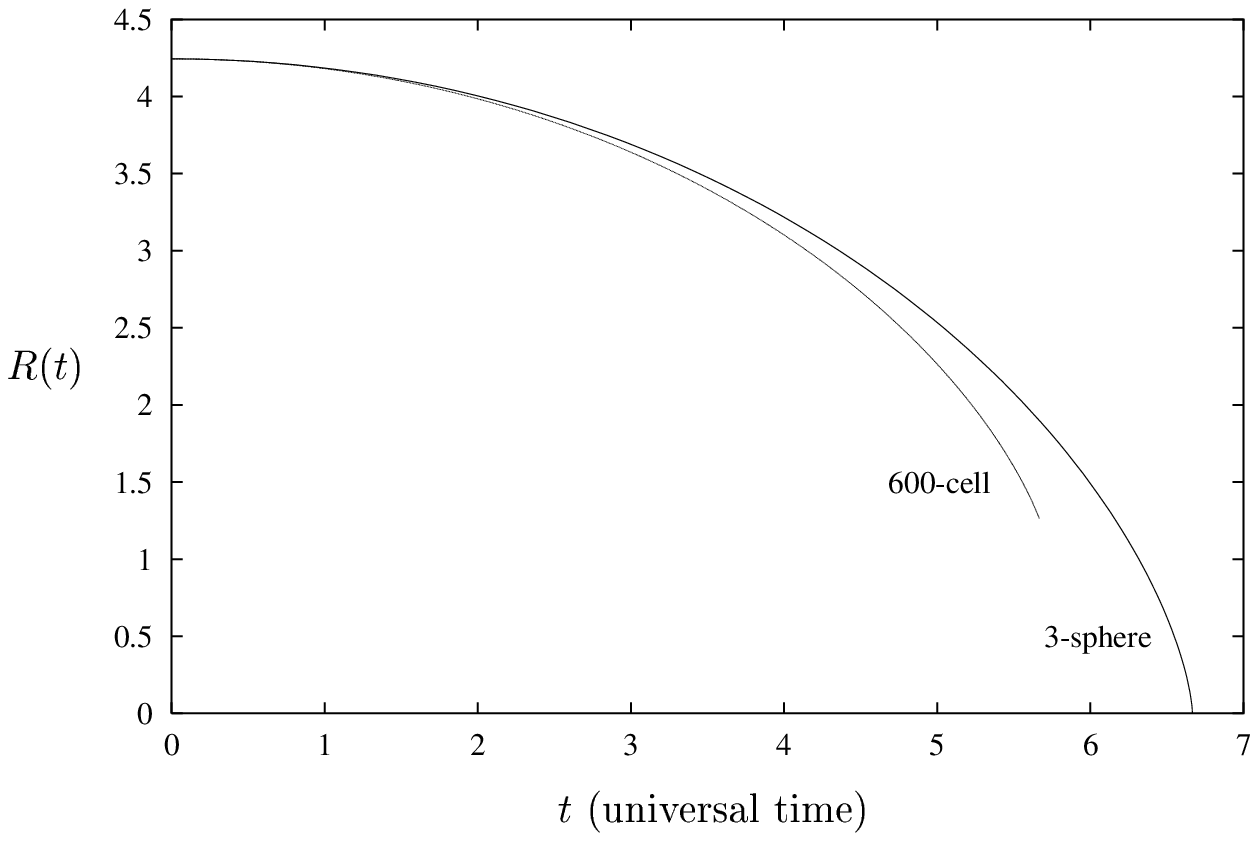}}\bigskip 
\figure{Evolution of the radius $R$ as a function of the universal time $t$
through \CdR\ and \RG\label{rfunzionet}}

\vfill\eject\null\bigskip

\centerline{\epsfbox{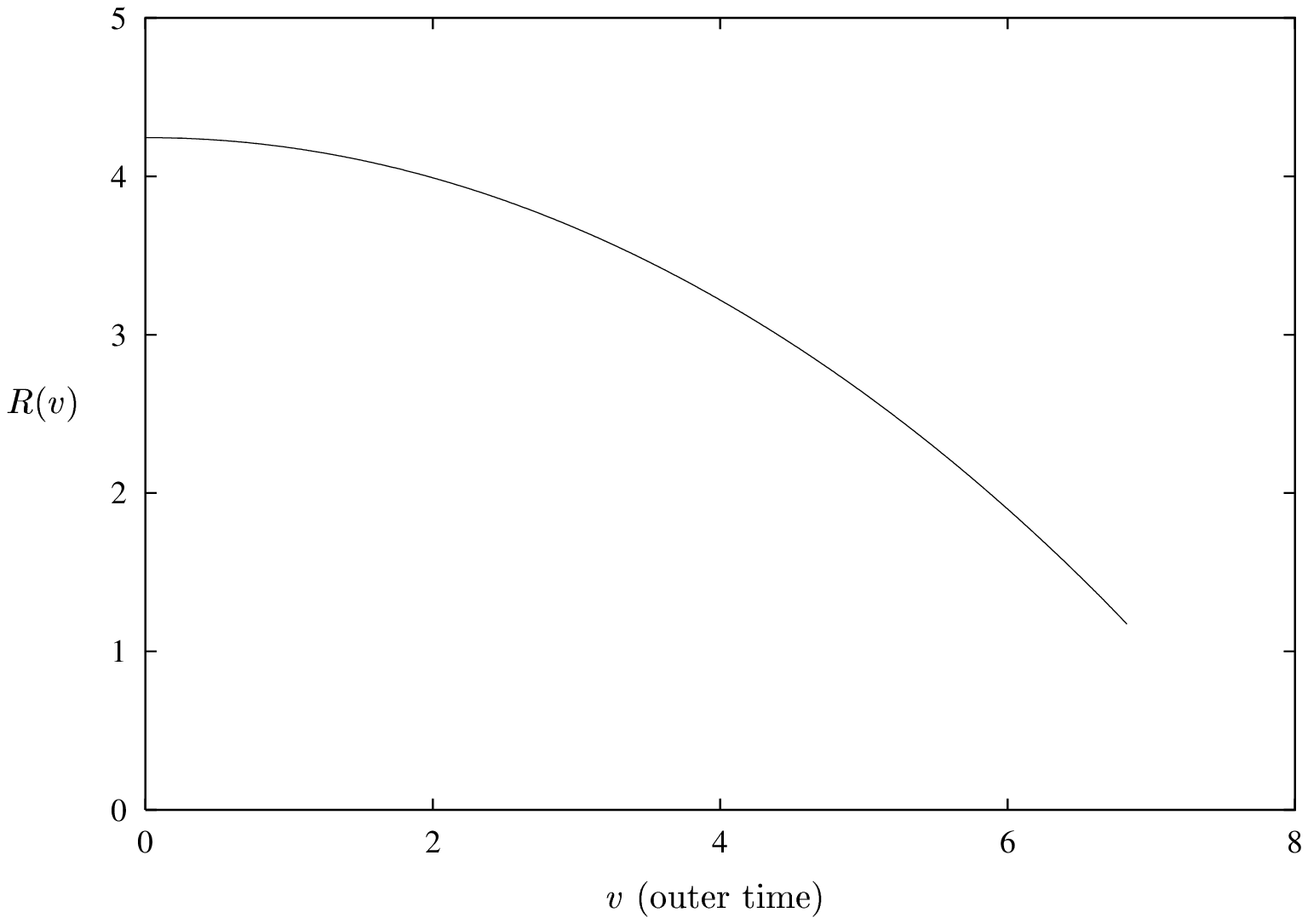}}\bigskip 
\figure{The behaviour of $R$ as a function of $v$.\label{rfunzionev}}

\vfill\eject\null\bigskip

\centerline{\epsfbox{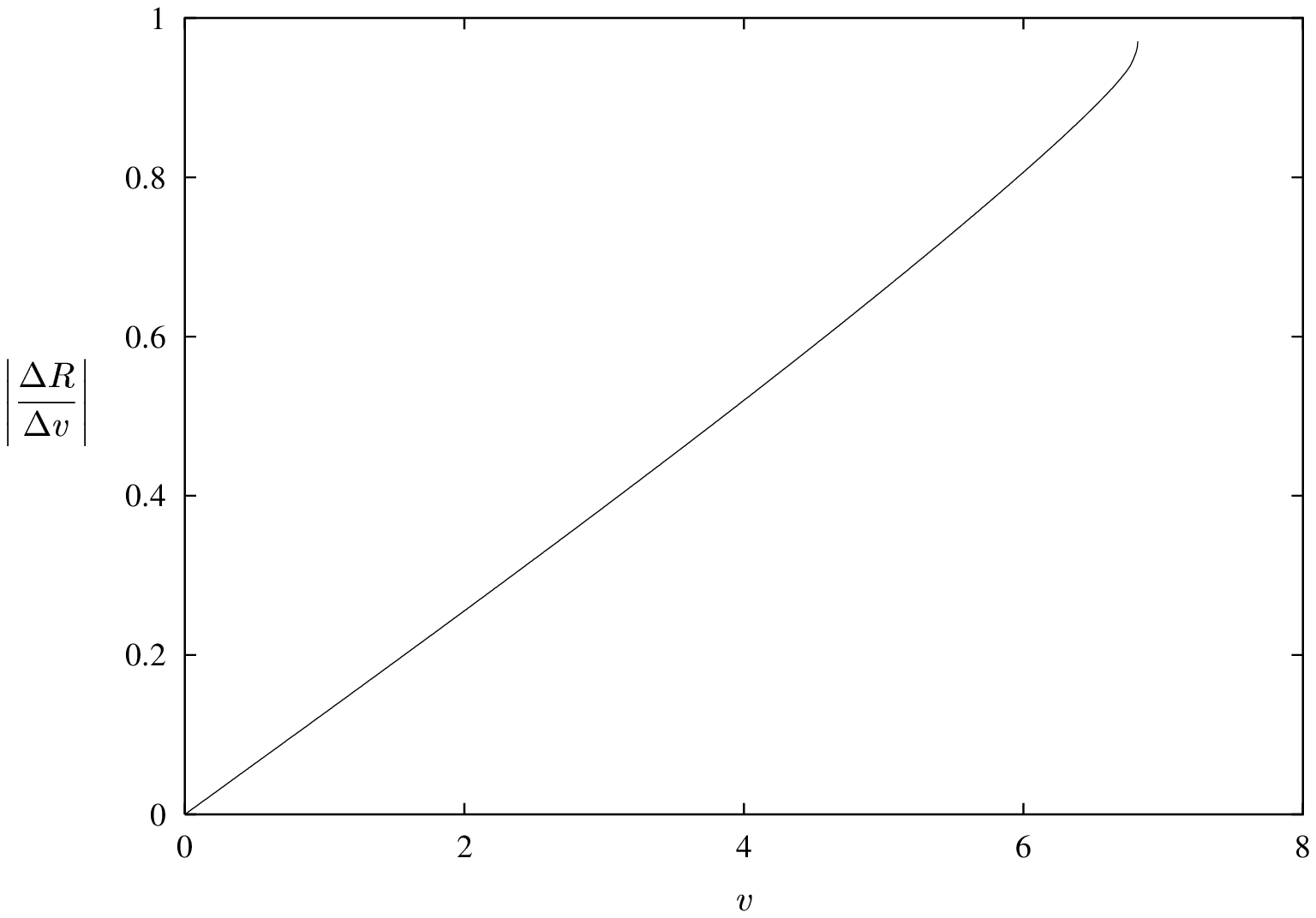}}\bigskip 
\figure{The behaviour of $|\Delta R/\Delta v|\approx \valass{\dot R}$ as a
function of $v$ for the 600-cell\label{dvfunzionedr}}
\bye